\documentclass[graybox, natbib, footinfo]{svmult}

\usepackage{mathptmx}       
\usepackage{helvet}         
\usepackage{courier}        
\usepackage{type1cm}        
\usepackage{makeidx}         
\usepackage{graphicx}        
\usepackage{multicol}        
\usepackage[bottom]{footmisc}

\makeindex             

\begin{document}

\title*{Spectroscopic constraints for low-mass asteroseismic targets}
\author{Thierry Morel}
\institute{Thierry Morel \at Institut d'Astrophysique et de G\'eophysique, Universit\'e de Li\`ege, All\'ee du 6 Ao\^ut, B\^at. B5c, 4000 Li\`ege, Belgium, \email{morel@astro.ulg.ac.be}}

\maketitle

\abstract{A full exploitation of the observations provided by the CoRoT and {\it Kepler} missions depends on our ability to complement these data with accurate effective temperatures and chemical abundances. We review in this contribution the major efforts that have been undertaken to characterise late-type, seismic targets based on spectra gathered as part of the ground-based, follow-up campaigns. A specific feature of the spectroscopic studies of these stars is that the gravity can be advantageously fixed to the more accurate value derived from the pulsation spectrum. We describe the impact that such an approach has on the estimation of $T_{\rm eff}$ and [Fe/H]. The relevance of red-giant seismic targets for studies of internal mixing processes and stellar populations in our Galaxy is also briefly discussed.}

\section{Introduction}
\label{sect_introduction}
The great potential of asteroseismology to address some unresolved issues in stellar physics and even, as was discussed during this meeting, to study the stellar populations making up our Galaxy cannot be overstated. Yet these expectations cannot be completely met if some fundamental quantities that are not encoded in seismic data are not accurately known \citep[e.g.,][]{creevey12}. For this reason, by providing the effective temperature and chemical composition (but also other important information such as the $v\sin i$ or the binary status), a traditional field such as stellar spectroscopy will still play an important role in the future for the study of seismic targets. Conversely, asteroseismology can provide the fundamental quantities (e.g., mass, age, evolutionary status in the case of red giants) that are needed to best interpret the abundance data. These two fields are therefore closely connected and can greatly benefit from each other. 

The large discrepancies between the $\log g$ and [Fe/H] values derived from spectroscopy and those in the {\it Kepler} Input Catalog \citep{bruntt12,thygesen12} illustrate the clear superiority of spectroscopic techniques over photometric ones for the estimation of these two parameters. Determining accurate temperatures from photometric indices is also challenging in the presence of a significant (and patchy) reddening (e.g., for some CoRoT fields that lie close to the Galactic plane).

\section{The samples discussed}
\label{sect_samples}
Numerous spectroscopic analyses of individual seismic targets have been conducted during the last few years \citep[e.g.,][]{mathur13,morel13}. However, we will restrict ourselves here to discussing the results of studies dealing with a sizeable number of stars observed by either the CoRoT or the {\it Kepler} space missions.  

The CoRoT satellite operated either through the seismology (observations of a limited number of bright stars in the context of seismic studies) or the exoplanet (observations of numerous faint stars to detect planetary transits) channel. The parameters of a large number of stars in various evolutionary stages in the CoRoT exofields have been determined using an automated pipeline by \citet{gazzano10}, while a more comprehensive analysis of 19 red giants in the seismology fields has been presented by \citet{morel14}.\footnote{Note that the sample of \citet{morel14} discussed in the following contains a few stars which were eventually not observed by the satellite, as well as a number of benchmark stars used for validation purposes.} In the latter case, a standard analysis is employed that imposes excitation and ionisation equilibrium of iron based on the equivalent widths of a set of Fe I and Fe II lines. 

On the other hand, a study of dwarfs and giants in the {\it Kepler} field has been performed by \citet{bruntt12} and \citet{thygesen12}, respectively (the latter study superseding that of \citealt{bruntt11}). In both cases, the analysis relied on the spectral-synthesis software package {\tt VWA} \citep[see, e.g.,][]{bruntt02}.

Table~\ref{tab_uncertainties} gives for all the studies mentioned above the uncertainties associated to the determination of the parameters. Based on the (sometimes rather scanty) information provided in these papers, it may be concluded that these figures are claimed to be representative of the {\it accuracy} of the results. Although these measurements also suffer from limitations (e.g., calibration issues, angular diameter corrections, reddening), the satisfactory agreement with the less model-dependent estimates provided by interferometry for stars at near-solar metallicities \citep[e.g.,][]{bruntt10,huber12,morel14} suggests that the values quoted in Table~\ref{tab_uncertainties} for $T_{\rm eff}$ are reasonable in this metallicity regime (however, this may not be true for metal-poor stars where non-LTE and 3D effects become important; \citealt{lind12,dobrovolskas13}). Much more extensive and stringent tests can be expected in the future thanks to the advent of new long-baseline interferometric facilities. A comparison for a subset of {\it Kepler} targets between the parameters obtained by \citet{bruntt12} and \citet{thygesen12}, and those derived by two other methods has recently been presented by \citet{molenda_zakowicz13}. For the reader interested in the differences arising from the use of different spectroscopic methods, see, e.g., \citet{gillon_magain06} and \citet{creevey12}. The impact of the neglect of non-LTE effects on the parameters inferred from excitation and ionisation balance of iron is discussed by, e.g., \citet{lind12} and \citet{bensby14}.

\begin{table}
\scriptsize
\caption{Typical 1-$\sigma$ uncertainty of the parameter determination for the seismic targets. When available, the second row gives for a given study the uncertainties in case the gravity is fixed to the seismic value (see Sect.~\ref{sect_adopting_seismic_logg}). References: [1] \citet{gazzano10}; [2] \citet{morel14}; [3] \citet{bruntt12}; [4] \citet{thygesen12}.}
\label{tab_uncertainties}       
\begin{tabular}{p{3.2cm}p{1.8cm}p{2.3cm}p{0.8cm}p{0.9cm}p{0.9cm}p{1.0cm}}
\hline\noalign{\smallskip}
Type of stars & Magnitude range & Type of data & $\sigma_{\rm \, T_{eff}}$ & $\sigma_{\rm \, \log g}$ & $\sigma_{\rm \, [Fe/H]}$ & Reference\\
\noalign{\smallskip}\svhline\noalign{\smallskip}
Stars in CoRoT exofields     & 12 $<$ $r'$ $<$ 16            & medium resolution$^a$ & 140 & 0.27 & 0.19 & 1\\
Giants in CoRoT seismofields &  6 $<$ $V$ $<$ 9              & high resolution       &  85 & 0.20 & 0.10 & 2\\
                             &                               &                       &  60 & 0.07 & 0.08 & 2\\
Dwarfs in {\it Kepler} field &  7 $<$ $V_{\rm \, T}$ $<$ 10.5 & high resolution       &  70 & 0.08 & ...  & 3\\
                             &                               &                       &  60 & 0.03 & 0.06 & 3\\  
Giants in {\it Kepler} field &  7 $<$ $V$ $<$ 12             & high resolution       &  80 & 0.20 & 0.15 & 4\\
\noalign{\smallskip}\hline\noalign{\smallskip}
\end{tabular}
$^a$ Also small wavelength coverage ($\sim$200 \AA).
\end{table}

\section{Adopting the seismic gravity in spectroscopic analyses}
\label{sect_adopting_seismic_logg}
As has been exhaustively discussed in the recent literature, $\log g$ can be estimated in various ways from seismic observables: either from a detailed modelling of the oscillation spectrum or from scaling relations/grid-based methods that make use of $\Delta \unu$ (the average large frequency separation) and $\unu_{\rm max}$ (the frequency corresponding to maximum oscillation power). A number of empirical tests \citep[e.g.,][and references therein]{chaplin13} indicate that such estimates are likely more accurate than those derived from spectroscopic methods (typically 0.05 vs 0.15-0.20 dex). There is therefore an advantage in fixing the gravity to the seismic value in spectroscopic analyses, as is indeed now routinely done \citep[e.g.,][]{huber13}.\footnote{The possibility of using an independent and more accurate gravity estimate is also shared by stars with planetary transits \citep[e.g.,][]{torres12}.} We will first discuss in the following the quantitative impact of adopting the seismic gravity on the determination of $T_{\rm eff}$ and [Fe/H], and then turn our attention to the issue of the best metallicity to adopt when such a hybrid approach is employed. 

\subsection{Impact on the determination of the other parameters}
\label{sect_impact_fixing_logg}
For the {\it Kepler} targets, there is a good level of agreement in a statistical sense between the spectroscopic and seismic gravities, with no evidence for global systematic offsets: $\langle$$\log g$ (spectroscopy) -- $\log g$ (seismology)$\rangle$ = +0.08$\pm$0.07 for dwarfs \citep{bruntt12} and --0.05$\pm$0.30 dex for giants \citep{thygesen12}, respectively. However, large differences can be found on a star-to-star basis (up to 0.7 dex). As shown by \citet{morel11}, even larger discrepancies are evident for the red giants studied by \citet{gazzano10} (see also discussion by \citealt{valentini13} who independently re-analysed these data and found a more satisfactory agreement, especially for spectra with a low signal-to-noise ratio). On the other hand, the values are identical within the errors for all the giants analysed by \citet{morel14}.

\begin{figure}
\centering
\includegraphics[scale=.55, trim = 10mm 60mm 10mm 85mm, clip]{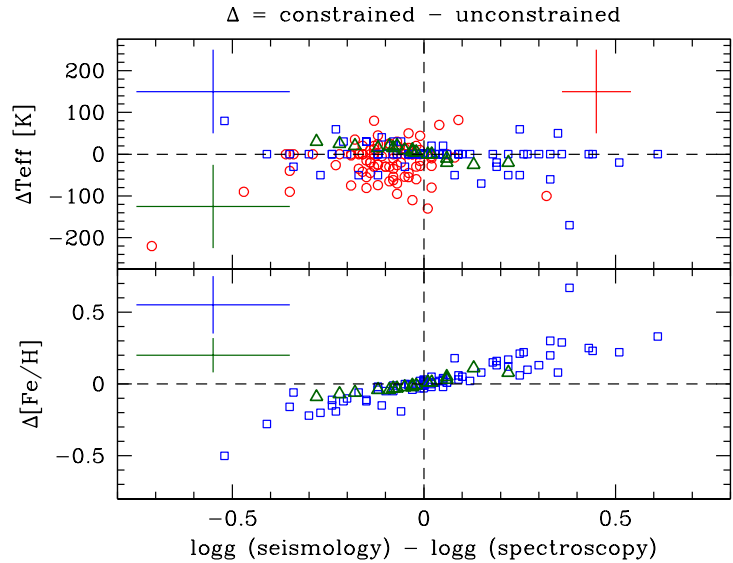}
\caption{Effect on the $T_{\rm eff}$ and [Fe/H] determinations of using either the seismic or the spectroscopic gravity. The (un)constrained results are for the $\log g$ (not) fixed to the seismic value. Red circles: {\it Kepler} dwarfs \citep{bruntt12}, blue squares: {\it Kepler} giants \citep{thygesen12}, green triangles: red giants in CoRoT seismofields \citep{morel14}. Note that the metallicities obtained using the spectroscopic gravities are not available for the {\it Kepler} dwarfs \citep{bruntt12}. The extreme outlier KIC~4070746 is not included in this figure \citep[see discussion in][]{thygesen12}.}
\label{fig_impact_adopting_seismic_logg}       
\end{figure}

The effect of fixing the gravity to the seismic value on the $T_{\rm eff}$ and [Fe/H] determinations is illustrated in Fig.~\ref{fig_impact_adopting_seismic_logg}. A change in $\log g$ of 0.1 dex typically leads for giants in the CoRoT seismology fields to variations in $T_{\rm eff}$ of 15 K and in [Fe/H] of 0.04 dex. The good agreement between the two sets of $\log g$ values only implies relatively small adjustments for $T_{\rm eff}$ and the abundances (generally below 50 K and 0.1 dex). A similar sensitivity of [Fe/H] against changes in $\log g$ is obtained for {\it Kepler} giants. However, variations in the adopted $\log g$ are in this case not accompanied in a coherent way by $T_{\rm eff}$ changes. It is in particular not completely clear how $\log g$ changes amounting to up to 0.6 dex can lead to exactly identical $T_{\rm eff}$ values. There is also a lack of correlation between the $\log g$ and $T_{\rm eff}$ changes for {\it Kepler} dwarfs. On the other hand, \citet{huber13} found for exoplanet host candidates (mostly solar-like) that a change in $\log g$ of 0.1 dex typically leads to variations of 50 K and 0.03 dex for $T_{\rm eff}$ and [Fe/H], respectively. It is important to note that the figures quoted above for dwarfs and giants cannot be generalised and depend on the exact procedures that are implemented to derive the parameters \citep[see][]{torres12}. 

\subsection{The ambiguity surrounding the best metallicity value} 
\label{sect_ambiguity_metallicity}
The surface gravity is usually determined from spectroscopic data by requiring that ionisation balance of iron is fulfilled. In many cases, this condition will no longer be satisfied once the seismic gravity is adopted \citep[][]{bruntt12,thygesen12}. As a result, the mean abundances derived from the Fe I and Fe II lines will differ, and there will therefore be an ambiguity as to which iron abundance should be preferred. As an illustration, using the seismic constraints, \citet{bruntt12} obtained [Fe I/H] = --0.02 and [Fe II/H] = +0.32 for KIC~3424541. The metallicity is an essential ingredient of any seismic modelling, and adopting one value or the other will clearly lead to substantially different estimates for the fundamental stellar parameters, such as the age, for instance. 

The Fe II lines are known in solar-like dwarfs to be less affected than the Fe I lines by both non-LTE and granulation effects \citep[e.g.,][]{asplund00}. The mean Fe II-based abundance hence appears to be  a better proxy of the stellar metallicity when using a 1D LTE analysis. However, the choice is not as straightforward for red giants. Although the departures from LTE are also much less severe for the Fe II lines, these features are affected by a number of caveats: (1) they are only usually a few, difficult to measure, and potentially more affected by blends; (2) they are very sensitive to errors in the effective temperature (varying $T_{\rm eff}$ by 50 K while keeping the gravity fixed typically changes the Fe I abundances by only 0.01-0.02 dex, but the Fe II ones by 0.06 dex); (3) they may suffer more than the Fe I lines at near-solar metallicity from the neglect of granulation effects (\citealt{collet07}; \citealt{kucinskas13}; see also fig.15 of \citealt{dobrovolskas13}). In view of the uncertainties plaguing both the Fe I and Fe II abundances, it is unclear whether the Fe II-based abundances should be deemed as (systematically) more reliable for evolved objects.

\section{A step beyond the determination of the basic parameters: the detailed chemical composition}
\label{sect_abundances}
\begin{figure}
\centering
\includegraphics[scale=.62, trim = 10mm 68mm 10mm 68mm, clip]{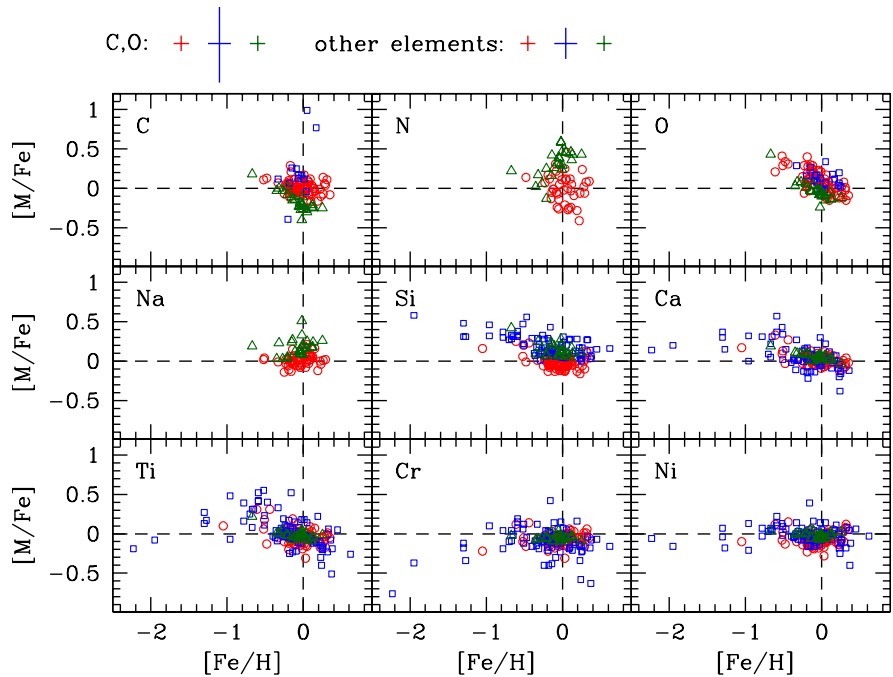}
\caption{Abundance ratios with respect to iron as a function of [Fe/H] for stars in the {\it Kepler} and CoRoT fields. The results have been obtained using a 1D LTE analysis and (except for the CoRoT stars) the seismic gravities. Same symbols as in Fig.~\ref{fig_impact_adopting_seismic_logg}. For the {\it Kepler} stars, [Fe/H] is based on the Fe II lines and the abundances of the other elements on the neutral species. Following \citet{bruntt12}, we only consider mean abundances for {\it Kepler} dwarfs with $v\sin i$ below 25 km s$^{-1}$ and computed based on at least five lines of each element (except for nitrogen and oxygen: 2 and 3 lines, respectively).}
\label{fig_abundances_vs_Fe}       
\end{figure}

The detailed abundance pattern can be obtained for stars observed with high-resolution spectrographs. Figure \ref{fig_abundances_vs_Fe} shows some abundance ratios with respect to iron as a function of [Fe/H] for the samples of \citet{bruntt12}, \citet{thygesen12}, and \citet{morel14}. Because of the chemical evolution of the Galaxy, it is well established that - depending on their nucleosynthesis - each element displays a distinct behaviour as a function of the iron content. For instance, the abundance ratio of the $\ualpha$ elements (e.g., Ca) increases when [Fe/H] decreases, whereas the iron-peak elements (e.g., Ni) closely follow Fe. An enhancement with respect to solar of some important species such as oxygen should be taken into account when modelling low-metallicity asteroseismic targets. Some elements behave qualitatively as expected in Fig.~\ref{fig_abundances_vs_Fe} (e.g., Si and Ni), but the expected trends at low metallicities are not seen in some cases (e.g., Ti and Cr) and the patterns generally much noisier than those reported in the literature for disc stars in the solar neighbourhood \citep[e.g.,][]{bensby14}. This can be at least partly attributed to the limitations of (semi)automated pipelines applied to data of lower quality. The fainter {\it Kepler} targets have often only been observed with 1m- or 2m-class telescopes \citep{bruntt12,thygesen12,molenda_zakowicz13}.

The data shown in Fig.~\ref{fig_abundances_vs_Fe} are heterogeneous and any study-to-study difference in the global patterns may be misinterpreted as being of physical origin whereas it merely reflects systematic effects. However, the carbon depletion and nitrogen excess of the CoRoT giants compared to {\it Kepler} dwarfs may be expected because of mixing \citep[see, e.g.,][in the case of C]{luck_heiter07}. More robust conclusions could have been drawn for carbon by comparing the data for {\it Kepler} dwarfs and giants (thanks to the similarity of the analyses carried out by \citealt{bruntt12} and \citealt{thygesen12}), but the results for giants are affected by large uncertainties.

\begin{figure}[h]
\sidecaption
\includegraphics[scale=.59, trim = 45mm 155mm 55mm 35mm, clip]{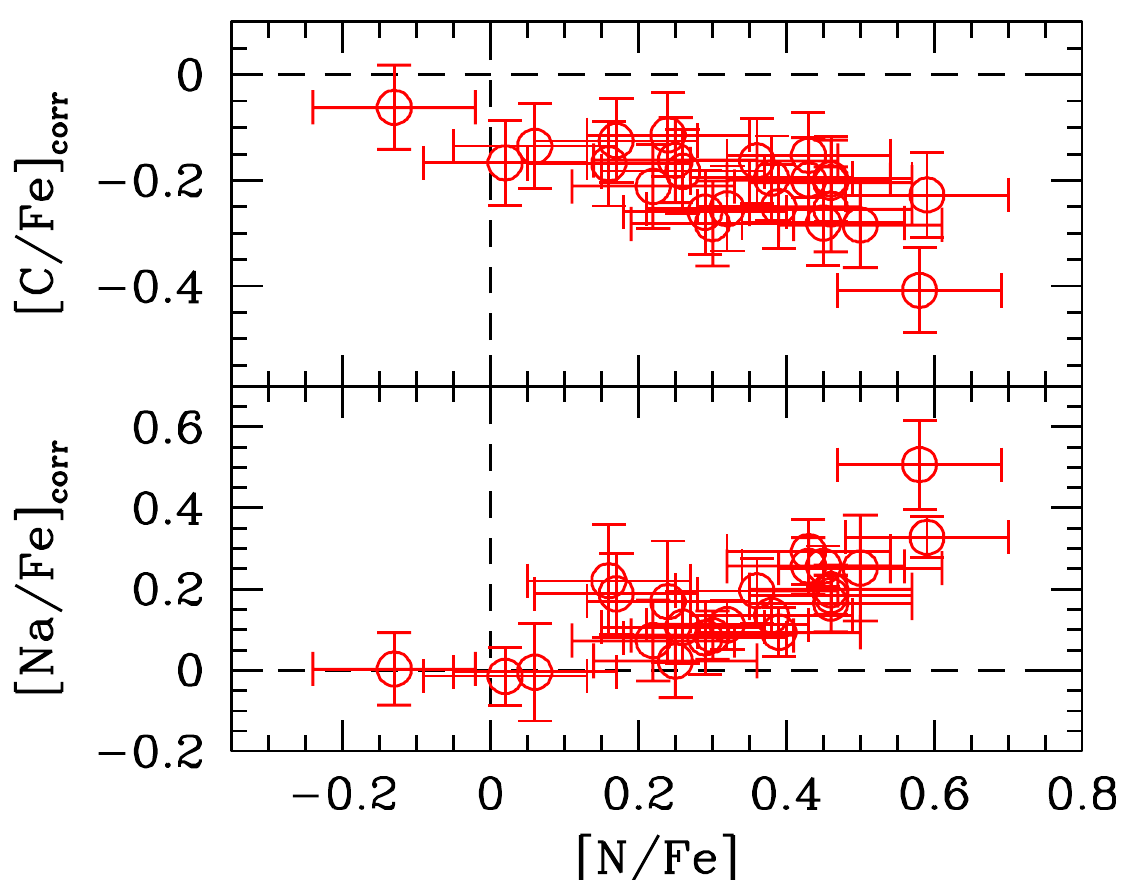}
\caption{Top and bottom panels: [C/Fe] and [Na/Fe] as a function of [N/Fe] for the red giants in the CoRoT seismology fields. The C and Na data have been corrected for the effects of the chemical evolution of the Galaxy \citep[for details, see][]{morel14}. The results have been obtained using the spectroscopic gravities.}
\label{fig_C_Na_vs_N}       
\end{figure}

The extent of mixing experienced by red giants results from the combined action of different physical processes (convective and rotational mixing, as well as arguably thermohaline instabilities) whose relative efficiency is a complex function of their evolutionary status, mass, metallicity, and rotational history \cite[e.g.,][]{charbonnel_lagarde10}. Fortunately, several key indicators with a different sensitivity to each of these processes can be measured in the optical wavelength domain (Li, CNO, Na, and $^{12}$C/$^{13}$C) and used to constrain theoretical models. As can be seen in Fig.~\ref{fig_C_Na_vs_N}, the occurrence of internal mixing phenomena is betrayed in CoRoT red giants by the existence of well-defined trends between the surface abundances of some species \citep[for a discussion of these results, see][]{morel14}. It is important to note that such abundance studies of asteroseismic targets may lead to a leap forward in our understanding of transport phenomena in evolved, low- and intermediate-mass stars because of the availability of an accurate mass estimate and, in some cases, a knowledge of the evolutionary status. 

\section{Some perspectives}
\label{sect_perspectives}
A detailed spectroscopic analysis has so far been carried out for only a tiny fraction of all the stars observed by CoRoT and {\it Kepler}. Much more is expected (or may be achievable) in the near future. We briefly mention below two of the most promising avenues of research.
 
Seismic targets are currently used as benchmark stars in various ongoing or soon-to-be-started large-scale surveys, such as APOGEE \citep{meszaros13}, Gaia-ESO \citep{gilmore12}, or GALAH \citep{freeman12}. The combination of these spectroscopic data with the asteroseismic ones for the radii, masses, ages, and distances will be of great relevance for investigating the properties of the stellar populations constituting our Galaxy \cite[see, e.g.,][]{chiappini12,miglio13}. The {\it Gaia} satellite will dramatically contribute to this harvest by providing kinematic information of unprecedented quality. 

The various evolutionary sequences of red-giant stars can be distinguished from asteroseismic diagnostics \citep[e.g.,][]{stello13,montalban13}. This opens up the possibility of mapping out the evolution of the mixing indicators during the shell-hydrogen and core-helium burning phases for a very large number of stars with accurate masses \citep[see the tentative results for carbon of][]{luck_heiter07}. Knowing the fundamental parameters (e.g., mass, age) of dwarfs and having the possibility of probing their internal structure also make them particularly suitable for investigating the destruction of lithium during the early stages of stellar evolution.

\begin{acknowledgement}
I acknowledge financial support from Belspo for contract PRODEX GAIA-DPAC. I am very grateful to the Fonds National de la Recherche Scientifique (FNRS) and Annie Baglin for providing the financial resources that made my attendance possible. 
\end{acknowledgement}

\bibliographystyle{spphys}
\bibliography{morel_sesto}

\begin{thebibliography}{1}
\providecommand{\url}[1]{{#1}}
\providecommand{\urlprefix}{URL }
\expandafter\ifx\csname urlstyle\endcsname\relax
  \providecommand{\doi}[1]{DOI \discretionary{}{}{}#1}\else
  \providecommand{\doi}{DOI \discretionary{}{}{}\begingroup
  \urlstyle{rm}\Url}\fi

\bibitem[Asplund et al.(2000)]{asplund00} Asplund, M., Nordlund, \AA, Trampedach, R., Stein, R. F.: \aap {\bf 359}, 743 (2000)
\bibitem[Bensby et al.(2014)]{bensby14} Bensby, T., Feltzing, S., Oey, M. S.: \aap, {\bf 562}, A71 (2014)
\bibitem[Bruntt et al.(2002)]{bruntt02} Bruntt, H., Catala, C., Garrido, R., et al.: \aap {\bf 389}, 345 (2002)
\bibitem[Bruntt et al.(2010)]{bruntt10} Bruntt, H., Bedding, T. R., Quirion, P.-O., et al.: \mnras {\bf 405}, 1907 (2010)
\bibitem[Bruntt et al.(2011)]{bruntt11} Bruntt, H., Frandsen, S., Thygesen, A. O.: \aap {\bf 528}, A121 (2011)
\bibitem[Bruntt et al.(2012)]{bruntt12} Bruntt, H., Basu, S., Smalley, B., et al.: \mnras {\bf 423}, 122 (2012)
\bibitem[Chaplin \& Miglio(2013)]{chaplin13} Chaplin, W. J., Miglio, A.: \araa {\bf 51}, 353 (2013)
\bibitem[Charbonnel \& Lagarde(2010)]{charbonnel_lagarde10} Charbonnel, C., Lagarde, N.: \aap {\bf 522}, A10 (2010)
\bibitem[Chiappini(2012)]{chiappini12} Chiappini, C.: in ``Red Giants as Probes of the Structure and Evolution of the Milky Way'', \apss Proc., 147 (2012)
\bibitem[Collet et al.(2007)]{collet07} Collet, R., Asplund, M., Trampedach, R.: \aap {\bf 469}, 687 (2007)
\bibitem[Creevey et al.(2012)]{creevey12} Creevey, O. L., Do\u{g}an, G., Frasca, A., et al.: \aap {\bf 537}, A111 (2012)
\bibitem[Dobrovolskas et al.(2013)]{dobrovolskas13} Dobrovolskas, V., Ku\v{c}inskas, A., Steffen, M., et al.: \aap  {\bf 559}, A102 (2013)
\bibitem[Freeman(2012)]{freeman12} Freeman, K. C.: ASPC {\bf 458}, 393 (2012)
\bibitem[Gazzano et al.(2010)]{gazzano10} Gazzano, J.-C., de Laverny, P., Deleuil, M., et al.: \aap {\bf 523}, A91 (2010)
\bibitem[Gillon \& Magain(2006)]{gillon_magain06} Gillon, M., Magain, P.: \aap {\bf 448}, 341 (2006) 
\bibitem[Gilmore et al.(2012)]{gilmore12} Gilmore, G., Randich, S., Asplund, M., et al.: The Messenger {\bf 147}, 25 (2012)
\bibitem[Huber et al.(2012)]{huber12} Huber, D., Ireland, M. J., Bedding, T. R., et al.: \apj {\bf 760}, 32 (2012)
\bibitem[Huber et al.(2013)]{huber13} Huber, D., Chaplin, W. J., Christensen-Dalsgaard, J., et al.: \apj {\bf 767}, 127 (2013)
\bibitem[Ku\v{c}inskas et al.(2013)]{kucinskas13} Ku\v{c}inskas, A., Steffen, M., Ludwig, H.-G., et al.: \aap {\bf 549}, A14 (2013)
\bibitem[Lind et al.(2012)]{lind12} Lind, K., Bergemann, M., Asplund, M.: \mnras {\bf 427}, 50 (2012)
\bibitem[Luck \& Heiter(2007)]{luck_heiter07} Luck, R. E., Heiter, U.: \aj {\bf 133}, 2464 (2007)
\bibitem[Mathur et al.(2013)]{mathur13} Mathur, S., Bruntt, H., Catala, C., et al.: \aap {\bf 549}, A12 (2013)
\bibitem[M\'esz\'aros et al.(2013)]{meszaros13} M\'esz\'aros, S., Holtzman, J., Garc\'{\i}a P\'erez, A. E., et al.: \aj {\bf 146}, 133 (2013)
\bibitem[Miglio et al.(2013)]{miglio13} Miglio, A., Chiappini, C., Morel, T., et al.: \mnras {\bf 429}, 423 (2013)
\bibitem[Molenda-\.Zakowicz et al.(2013)]{molenda_zakowicz13} Molenda-\.Zakowicz, J., Sousa, S. G., Frasca, A., et al.: \mnras {\bf 434}, 1422 (2013)
\bibitem[Montalb\'an et al.(2013)]{montalban13} Montalb\'an, J., Miglio, A., Noels, A., et al.: \apj {\bf 766}, 118 (2013)
\bibitem[Morel et al.(2011)]{morel11} Morel, T., Miglio, A., Valentini, M.: JPhCS {\bf 328}, 012010 (2011)
\bibitem[Morel et al.(2013)]{morel13} Morel, T., Rainer, M., Poretti, E., Barban, C., Boumier, P.: \aap {\bf 552}, A42 (2013)
\bibitem[Morel et al.(2014)]{morel14} Morel, T., Miglio, A., Lagarde, N., et al.: \aap, in press (arXiv:1403.4373) (2014)
\bibitem[Stello et al.(2013)]{stello13} Stello, D., Huber, D., Bedding, T. R., et al.: \apjl {\bf 765}, L41 (2013)
\bibitem[Thygesen et al.(2012)]{thygesen12} Thygesen, A. O., Frandsen, S., Bruntt, H., et al.: \aap {\bf 543}, A160 (2012)
\bibitem[Torres et al.(2012)]{torres12} Torres, G., Fischer, D. A., Sozzetti, A., et al.: \apj {\bf 757}, 161 (2012)
\bibitem[Valentini et al.(2013)]{valentini13} Valentini, M., Morel, T., Miglio, A., Fossati, L., Munari, U.: EPJWC {\bf 43}, 03006 (2013)

\end{thebibliography}

\end{document}